\documentclass[letterpaper,fleqn,usenatbib]{mnras}

\def\cf{{\it cf.}}
\def\eg{{\it e.g.}}
\def\etal{{\it et al.}}

\def\ie{{\it i.e.}}

\def\pmb#1{\setbox0=\hbox{$#1$}%
  \kern-0.25em\copy0\kern-\wd0
  \kern.05em\copy0\kern-\wd0
  \kern-0.025em\raise.0433em\box0}
\def\spmb#1{\setbox1=\hbox{${\scriptstyle #1}$}%
  \kern-0.25em\copy1\kern-\wd1
  \kern.05em\copy1\kern-\wd1
  \kern-0.025em\raise.0433em\box1}

\long\def\Ignore#1{\relax}
\usepackage{color}
\definecolor{red}{rgb}{0.7,0.1,0.1}
\definecolor{blue}{rgb}{0.2,0.2,0.8}
\definecolor{green}{rgb}{0.1,0.6,0.1}

\usepackage{graphicx}	

\title[Responses to Noise II]{Spiral Instabilities in $N$-body Simulations: Emergence from noise II}

\author[Sellwood]
          {J. A. Sellwood,$^{1}$\thanks{E-mail:sellwood@as.arizona.edu}
\\
$^1$Steward Observatory, University of Arizona, 933 N Cherry Ave,
Tucson AZ 85722, USA}

\pubyear{2019}

\begin{document}
\label{firstpage}
\pagerange{\pageref{firstpage}--\pageref{lastpage}}
\maketitle

\begin{abstract}
An earlier paper presented the potentially significant discovery that
disturbances in simplified simulations of a stellar disc model that
was predicted to be stable in linear theory grew to large amplitude
over a long period of time.  The ultimate appearance of true
instabilities was attributed to non-linear scattering by a succession
of collective waves excited by shot noise from the finite number of
particles.  The paper concluded that no finite number of particles,
however large, could mimic a smooth disc.  As this surprising finding
has been challenged as an artifact of the numerical scheme employed,
we here present a new calculation of the same model using a different
grid geometry that confirms the original behaviour.
\end{abstract}



\section{Introduction}
\label{sec.intro}
\citet[][following Zang 1976]{To81} made the remarkable prediction
that the half-mass Mestel disc, whose properties are described below
(\S\ref{sec.methods}), is globally stable to gravitationally-driven
disturbances in the stellar distribution.  His prediction was based on
the usual linear stability analysis of small-amplitude perturbations
in which second-order terms are neglected \citep[\eg][]{Ka71, BT08}.
Toomre's prediction is important since it remains the most
interesting, smooth disc model to be predicted to stable.\footnote{Two
  stable ``composite'' disc models were also presented by
  \citet{Ka72}, but uniformly rotating discs are implausible models of
  galaxies and are more easily stabilized because they do not support
  swing amplification.}  Furthermore, a stable disk model enables
phenomena such as swing-amplification to be studied in a global
context \citep{To81}, and can be modified to excite instabilities in a
controlled manner \citep{SK91, SB02, SC19}.

\citet[][hereafter S12]{Se12} presented simulations to test Toomre's
prediction that the unmodified Mestel disc should be stable.  The
initial behaviour of his larger $N$ simulations (up to $N=5 \times
10^8$) broadly confirmed Toomre's prediction, but S12 also reported
that all his $N$-body realizations of this model were ultimately
non-linearly unstable.  The interesting new phenomenon that eventually
dominated his long-duration simulations was caused by the response of
the disc to shot noise from the finite number of particles.  He showed
\citep[see also Figure 1b of][]{SC19} that multiple episodes of
swing-amplified shot noise each launched inwardly travelling
disturbances that were absorbed at their inner Lindblad resonances.
The non-linear scattering of particles as they absorbed energy from
each wave, a consequence that is neglected in all linear stability
analyses, created a ``scratch'' \cite[or ``scar'' \cf][]{Sr19} in the
distribution function that enhanced the responsiveness of the disc to
subsequent noise-driven disturbances.  The extra responsiveness caused
the amplitude of successive disturbances to increase slowly at first,
but eventually led to a full-blown linear instability.

\citet{FP15} and \citet{FBP15} showed that the mildly non-linear
behaviour reported by S12 was quantitatively consistent with the
predictions of diffusion coefficients computed from the
self-gravitating responses to shot noise.  Furthermore, linear
stability analysis by \citet{DFP19} of the modified $N=50$M particle
distribution confirmed the same unstable mode that S12 had fitted to
data from his simulations.

Because one should always be suspicious of surprising behaviour in
simulations, S12 conducted a painstaking search for a possible
numerical cause of the new instabilities.  But since all his tests
employed the same grid geometry, it remains possible that
some kind of artifact of the polar grid could be responsible for
misbehaviour.  Two examples of possible inadequacies are (1) that the
sizes of the grid cells on a polar grid increase linearly with radius,
causing grid resolution to degrade with increasing radius, and (2)
grid aliasing could cause a tightly wrapped trailing wave to seed a
tightly wrapped leading disturbance; were this to happen, it would be
a numerical cause of indefinite growth.  S12 was aware of these
possibilities, which prompted him to test the consequences of halving
the grid cell sizes repeatedly, only to find that neither the rate of
growth of non-axisymmetric disturbances nor the overall qualitative
behaviour depended on grid resolution, as might be expected if either
of these hypothesized causes were responsible for misbehaviour.
However, a more convincing test would be to check whether the
evolution is affected in any way by the choice of grid geometry, which
is the purpose of the present paper.  Here we show that new a
simulation of the same physical model that employed a Cartesian grid
behaves in a strikingly similar manner to the corresponding case
reported in S12, confirming that the original, somewhat surprising,
findings of that paper were not a numerical artifact of the polar
grid.

\section{Technique}
\label{sec.methods}
\subsection{Disc model}
The razor-thin Mestel disc used in the studies by \citet{Za76} and
\citet{To81} is characterized by a constant circular speed $V_0$ at
all radii.  The axisymmetric surface density $\Sigma_0(R) = V_0^2 /
(2\pi G R)$ would self-consistently yield the appropriate central
attraction for centrifugal balance.  The lop-sided instability of the
full mass disc found by \citet{Za76} is suppressed when the surface
density is halved, with the removed mass added to a rigid halo to
maintain centrifugal balance.  These authors employed the distribution
function (DF) given by \citet{To77} and \citet{BT08}, which for the
half-mass disk is
\begin{equation}
f(E,L_z) = 0.5 \cases{ F \, (L_z/R_0V_0)^q e^{-E/\sigma_R^2} & $L_z>0$ \cr
           0 & otherwise, \cr}
\label{eq.DF}
\end{equation}
where $R_0$ is a reference radius, $q = V_0^2/\sigma_R^2 - 1$, and the
normalization constant is
\begin{equation}
F = {1 \over G R_0} { (q/2 + 0.5)^{q/2+1} \over \pi^{3/2}(q/2-0.5)!}.
\end{equation}
Choosing $q=11.44$ yields a Gaussian distribution of velocities such
that the half-mass disc has $Q=1.5$.  Toomre further multiplied the DF
$f$ by the double taper function
\begin{equation}
T(L_z) = 
\left[ 1 + \left( {R_0V_0 \over L_z} \right)^\nu \right]^{-1}
\left[ 1 + \left( {L_z \over R_1V_0} \right)^\mu \right]^{-1},
\label{eq.tapers}
\end{equation}
to create a central cut out and an outer taper having mean radii $R_0$
and $R_1$ respectively, while maintaining the centripetal acceleration
$-V_0^2/R$ everywhere.  Setting the taper indices $\nu=4$ and $\mu=5$
yielded an idealized, smooth disc model that Toomre claimed possessed
no small amplitude unstable modes.  S12 chose $R_1 = 11.5R_0$, and
limited the radial extent of the disc by an energy cut-off that
eliminated particles having sufficient energy to pass $R=20R_0$.  As
usual, we adopt units such that $G=V_0=R_0=1$.

\subsection{Numerical method}
Since the behaviour in simulations with unrestricted forces can be
quite complicated, S12 simplified his experiments by including only
force terms that arose from bisymmetric disturbances in the particle
distribution.  As all stability analyses take advantage of the
separation of individual sectoral harmonics in the determination of
the gravitational field \citep{BT08}, Sellwood's simplification
mirrored that of the prediction he was attempting to verify.  The ease
with which restrictions to the range of sectoral harmonics that
contribute to the gravitational field can be accomplished on the polar
grid made it a natural choice for the calculations presented in S12.

S12 therefore employed a 2D polar grid having $N_R$ rings and $N_A$
spokes.  The grid rings were spaced as
\begin{equation}
R_g(u) = h_R \left(e^{\alpha u} - 1 \right),
\label{eq.grofu}
\end{equation}
with $\alpha = 2\pi/N_A$, $u$ taking integer values in the range $0
\leq u \leq N_R - 1$ so that $R_g(0) = 0$, and the quantity $h_R$
defines the spatial scale.  Thus the fixed angular separation of the
spokes, together with the almost linear increase in the radial spacing
of the grid rings, causes grid cells with $e^{\alpha u} \gg 1$ to have
similar shapes but to increase in size with radius.  S12 chose $h_R =
R_0/8$ and $(N_R,N_A) = (106,128)$ for most simulations.

\begin{table}
\caption{Numerical parameters}
\label{tab.params}
\begin{tabular}{@{}ll}
Cartesian grid size & 1024 $\times$ 1024 \\
$R_0$ & 25 grid units \\
Softening length & $R_0/8$ \\
Number of particles & $5 \times 10^7$ \\
Basic time-step & $0.025(R_0/V_0)$ \\
Time step zones & 5 \\
\end{tabular}
\end{table}

In order to test for possible misbehaviour that might have been caused
by that choice of grid, we here report fresh calculations of the same
physical model using a 2D Cartesian grid. The FFT method to determine
the gravitational acceleration, with grid doubling in both $x$ and
$y$, and the standard cloud-in-cell (\ie\ linear) interpolation scheme
is described in \citet{Se14}.  The numerical parameters of the two
simulations presented here are given in Table~\ref{tab.params}.  For
efficiency, the code employs block time steps, with the time-step
being doubled in each of a series of annular zones whose boundaries
lie at radii that are themselves successively increased by factors of
two, as appropriate for a flat rotation curve.

In order to check the results of S12, we must again restrict the
accelerations to those arising only from bisymmetric disturbances in
the mass distribution.  We adopted the following two stratagems to
accomplish this goal on the Cartesian grid.

First, we began by computing an exactly smooth mass distribution on
the grid, created by integrating the tapered DF (eqs.~\ref{eq.DF} \&
\ref{eq.tapers}) over all velocities.  The mass assigned to each grid
point is therefore free from all shot noise that would arise from any
finite number of particles.  A simulation with a finite number of
particles then proceeds by assigning their masses to the grid in the
usual manner at every step, after which we subtract the pre-calculated
smooth mass distribution.  Consequently, the gravitational field
determined from this residual mass distribution is that arising only
from particle shot noise and its associated collective response.
To ensure dynamical equilibrium, we added the analytic central
attraction of the Mestel disc model, $a_R = -V_0^2/R$, to these
grid-determined acceleration components for each particle at each
step, as was also done in S12.

Second, there is no simple way to restrict the disturbance forces to
be perfectly bi-symmetric, but an excellent approximation to this goal
is easy to accomplish.  Taking advantage of the 4-fold rotation
symmetry of the grid, we replaced the mass assigned to each grid point
$d(i,j)$ by the following average at every step
\begin{equation}
d_4(i,j) = {1 \over 4}\left[ d(i,j) - d(-j,i) + d(-i,-j) - d(j,-i) \right],
\label{eq.bisym}
\end{equation}
with the indices $(i,j)$ reckoned from the grid centre.  A simple
average $d_2(i,j) = {1 \over 2}\left[ d(i,j) + d(-i,-j) \right],$
would retain contributions from all even sectoral harmonics with $m
\ge 2$, and we have found that disturbances grow somewhat more rapidly
when the $m=4$ term in particular contributed to the disturbance
forces.  However, eq.~(\ref{eq.bisym}) also suppresses all $m=4j$
terms, with integer $j\ge1$.  The first surving term is bisymmetric,
and although all $m = 2 + 4j$ terms also contribute to the disturbance
forces, the half-mass Mestel disk can support only very mild responses
to disturbances having rotational symmetries $m\ge6$, since Toomre's
swing-amplification parameter $X = 4/m$ in the half-mass Mestel disc.

\section{Results}
\label{sec.results}
Starting from the same file of 50M initial particle coordinates that
was used for model 50 in S12, we have used a Cartesian grid, with the
modifications just described, to recompute the evolution to $t=1600$.
We denote this calculation as run 50C.  Note that both runs 50 and 50C
were long integrations; the orbit period at $R=10$, half-way out in
the disk, is $20\pi$ in these units.

\begin{figure}
\includegraphics[width=.94\hsize,angle=0]{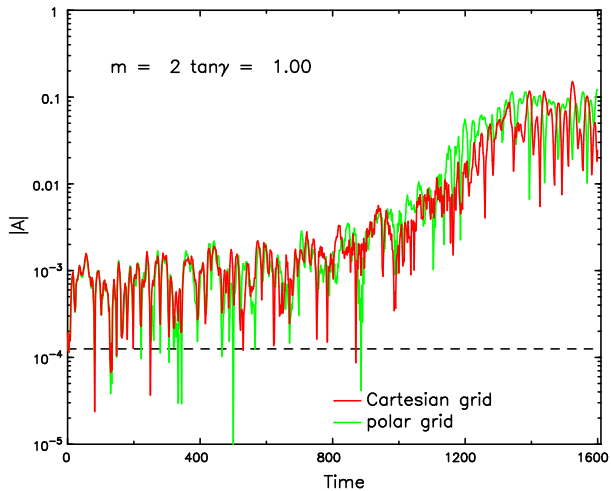}
\caption{The amplitude evolution of the given logarithmic spiral
  coefficient (eq.~\ref{eq.logspi}) in run 50C on the Cartesian grid
  (red curve) compared with the same quantity from run 50 reported in
  S12 (green curve). The dashed line indicates the expectation value
  $\langle|A|\rangle = (\pi/4N)^{1/2}$ if the particle distribution
  were random.}
\label{fig.gtest}
\end{figure}

Unlike with the polar grid, it is not straightforward to report the
amplitude of non-axisymmetric disturbances as function of radius in
simulations run on the Cartesian grid.  Thus we here employ a
different measure from that presented in S12: the amplitude of a
global logarithmic transform of the particle positions.  For equal
mass particles, the complex coefficients are evaluated as
\begin{equation}
A(m,\gamma,t) = {1 \over N} \sum_j e^{im(\phi_j + \tan\gamma \ln R_j)},
\label{eq.logspi}
\end{equation}
where $(R_j,\phi_j)$ are the polar coordinates of the $j$-th particle
at time $t$, and $\gamma$ is the angle between the tangent to an
$m$-arm logarithmic spiral and the radius vector, with positive values
for trailing spirals.

\begin{figure}
\includegraphics[width=.94\hsize,angle=0]{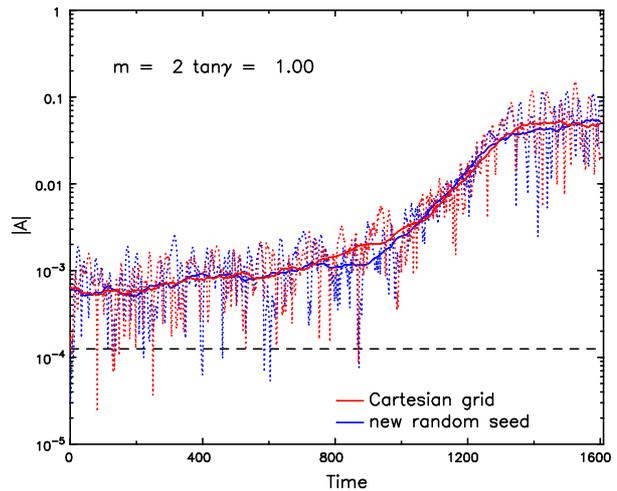}
\caption{The dotted curves are for run 50C (red), reproduced from
  Figure~\ref{fig.gtest}, and run 50R (blue) that started from a
  different random draw from the DF.  The solid curves are simple
  running averages.  Both simulations were run on the Cartesian grid
  with the parameters listed in Table~\ref{tab.params} and with
  disturbances forces restricted to $m=2 + 4j$ as described in the
  text.}
\label{fig.rantest}
\end{figure}

Figure~\ref{fig.gtest} compares the evolution of $|A(2,\pi/4,t)|$ from
run 50C (red curve) with the same quantity from the original run 50
(green curve) reported in S12, which employed the polar grid.  In both
cases, the dominant disturbance forces were restricted to $m=2$.  The
initial amplitude is consistent with random, as marked by the dashed
line, but rises by a factor of several in the first few dynamical
times as the gravitational forces polarize the particle distribution
\citep{TK91}.  Although this Figure is for $\tan\gamma=1$, implying a
trailing pitch angle of $45^\circ$, the agreement between the two
simulations is similarly impressive for $-1 \leq \tan\gamma \leq 3$.

In order to show the good agreement of even the high temporal
frequency amplitude variations, we have not attempted to smooth these
curves.  The rapid changes arise from the superposition of several
disturbances, each having different pattern speeds, which causes
alternating reinforcement and cancellation in the global transform
(\ref{eq.logspi}).  Since the two simulations began with the same file
of particle coordinates, they have the exact same initial noise
spectrum.  The two curves track each other perfectly in the early
stages, indicating identical dynamical responses to the noise.  Since
the system being simulated is stochastic \citep{SD09}, minor
differences emerge at later times that probably arise from the
cumulative effects of round-off error that must differ on the two
grids, but the overall evolution remains decisively similar.

Because all simulations employ approximations both to compute the
gravitational field and to integrate the motion of the particles,
none can be completely free from numerical artifacts.  But flaws
arising from the approximations associated with the calculation of the
field and interpolation between grid points, which differ
significantly between the two grid geometries, are most unlikely to
drive such similar evolution.

It is evident from Figure~\ref{fig.gtest} that the qualititative
evolution is unaffected by stochasticity, but in order to show that
there is nothing odd about the file of particles used for both models,
we present a second case, model 50R, that started from a different
random draw from the DF. (S12 reported that he had made a similar test
using the polar grid, but did not present the result.)  The amplitude
of the logarithmic spiral in run 50C (red line) is compared with that
from run 50R (blue line) in Figure~\ref{fig.rantest}, and in this
Figure we have also smoothed the data in order to emphasize the
agreement on longer time scales.

The overall behaviour of the disturbances in the three simulations 50,
50C and 50R is remarkably similar.  Aside from the high frequency
variations, all three runs manifest a gradually increasing amplitude
to $t \simeq 800$ followed by a more rapid increase, which peaks at $t
\simeq 1300$, by which time the inner disk is developing a bar in all
cases.  The close agreement at all times implies that the processes
of swing-amplification, resonance scattering, and enhanced
responsiveness identified by S12 occur to almost exactly the same
extent, and with the same time dependence, in all three of these
calculations.  Together with the dependence on particle number and the
other tests reported in S12, these new simulations strongly support
the earlier contention that the gradual emergence of instabilities is
physically real and does not result from inadequacies in the numerical
schemes.

\section{Conclusions}
\label{sec.concl}
The spectacular agreement, illustrated in Figure~\ref{fig.gtest},
between the result presented here and one of those reported by
\citet{Se12}, confirms that the secular rise in the amplitude of
disturbances does not result from artifacts of the grid used to
compute gravitational accelerations.  The earlier reported tests
involving variation of the time step, the number of time step zones,
locations of the zone boundaries, grid resolution, and random seed
(and also Figure~\ref{fig.rantest}) were all similarly reassuring.
These numerical checks leave little room to doubt the validity of the
results presented in S12.  The subsequent analyses by \citet{FP15},
\citet{FBP15}, \citet{DFP19} and \citet{Sr19} have also provided
welcome backup to the findings of S12.

Note that \citet{TK91} reported only ``an enhanced responsiveness''
and not a final runaway to instability in their long duration
simulations that employed the sheared sheet (local) approximation with
periodic boundary conditions.  However, in order to maintain long-term
responsiveness, those authors added a mild radial damping term to the
forces acting on the particles.  That seemingly innocent addition
almost certainly damped away the consequences of resonant scattering,
thereby preventing their simulations from manifesting any true
instabilities.  \citet{DVH13} employed $10^8$ particles in a global
simulation of an unperturbed low mass disc, concluding it was
``stable'' because no visible features appeared after ``two to three
galactic years.''  But their test was of too short duration for the
behaviour reported in S12 to have reached large amplitude.

The physical implication of the finding in S12, and confirmd here, is
that any differentially rotating disc of particles will eventually be
destabilized as a result of non-linear scattering by collective waves
seeded by shot noise, no matter how many particles are employed.

\section*{Acknowledgements}
The author acknowledges a vigorous correspondence with Agris Kalnajs,
which was the principal motivation for this work.  He is also grateful
to an anonymous referee for helpful comments, to Ray Carlberg for
advice, and to Steward Observatory for their continued hospitality.


\bsp	
\label{lastpage}

\begin{thebibliography}{99}
\def\skip#1{ \etal\ }
\def\PhD{PhD thesis.}
\def\rmp{Rev. Mod. Phys.}
\def\rpp{Rep. Prog. Phys.}

\bibitem[\protect\citeauthoryear{Binney \& Tremaine}{2008}]{BT08}
Binney J. \& Tremaine S. 2008, \textit{Galactic Dynamics} 2nd ed. (Princeton University Press, Princeton NJ) (BT08)

\bibitem[\protect\citeauthoryear{De Rijcke \etal}{2019}]{DFP19}
De Rijcke, S., Fouvry, J-B. \& Pichon, C. 2019, \mnras, {\bf 484}, 3198

\bibitem[\protect\citeauthoryear{D'Onghia \etal}{2013}]{DVH13}
D'Onghia, E., Vogelsberger, M. \& Hernquist, L. 2013, \apj, {\bf 766}, 34


\bibitem[\protect\citeauthoryear{Fouvry \etal}{2015}]{FBP15}
Fouvry, J-B., Binney, J. \& Pichon, C. 2015, \apj, {\bf 806}, 117

\bibitem[\protect\citeauthoryear{Fouvry \& Pichon}{2015}]{FP15}
Fouvry, J-B. \& Pichon, C. 2015, \mnras, {\bf 449}, 1928

\bibitem[\protect\citeauthoryear{Kalnajs}{1971}]{Ka71}
Kalnajs, A. J. 1971, \apj, {\bf 166}, 275

\bibitem[\protect\citeauthoryear{Kalnajs}{1972}]{Ka72}
Kalnajs, A. J. 1972, \apj, {\bf 175}, 63

\bibitem[\protect\citeauthoryear{Sellwood}{2012}]{Se12}
Sellwood, J. A. 2012, \apj, {\bf 751}, 44 (S12)

\bibitem[\protect\citeauthoryear{Sellwood}{2014}]{Se14}
Sellwood, J. A. 2014, arXiv:1406.6606 (on-line manual: \hfil\break {\tt http://www.physics.rutgers.edu/$\sim$sellwood/manual.pdf})

\bibitem[\protect\citeauthoryear{Sellwood \& Binney}{2002}]{SB02}
Sellwood, J. A. \& Binney, J. J. 2002, \mnras, {\bf 336}, 785

\bibitem[\protect\citeauthoryear{Sellwood \& Carlberg}{2019}]{SC19}
Sellwood, J. A. \& Carlberg, R. G. 2019, \mnras, {\bf 489}, 116

\bibitem[\protect\citeauthoryear{Sellwood \& Debattista}{2009}]{SD09}
Sellwood, J. A. \& Debattista, V. P. 2009, \mnras, {\bf 398}, 1279

\bibitem[\protect\citeauthoryear{Sellwood \& Kahn}{1991}]{SK91}
Sellwood, J. A. \& Kahn, F. D. 1991, \mnras, {\bf 250}, 278

\bibitem[\protect\citeauthoryear{Sridhar}{2019}]{Sr19}
Sridhar, S. 2019, \apj, {\bf 884}, 3

\bibitem[\protect\citeauthoryear{Toomre}{1977}]{To77}
Toomre, A. 1977, \araa, {\bf 15}, 437

\bibitem[\protect\citeauthoryear{Toomre}{1981}]{To81}
Toomre, A. 1981, In ''The Structure and Evolution of Normal Galaxies'', eds.~S. M. Fall \& D. Lynden-Bell (Cambridge, Cambridge Univ. Press) p.~111

\bibitem[\protect\citeauthoryear{Toomre \& Kalnajs}{1991}]{TK91}
Toomre, A. \& Kalnajs, A. J. 1991, in {\it Dynamics of Disc Galaxies}, ed.\ B. Sundelius (Gothenburg: G\"oteborgs University) p.~341

\bibitem[\protect\citeauthoryear{Zang}{1976}]{Za76}
Zang, T. A. 1976, \PhD, MIT

\end{thebibliography}
\end{document}